\documentclass[showpacs,aps,prd,reprint,groupedaddress,nofootinbib
]{revtex4-1}

\usepackage{amsmath,amssymb}
\usepackage{bm}
\usepackage{graphicx}
\usepackage{ascmac}
\usepackage{color}
\usepackage{longtable}
\usepackage{mathrsfs}

\usepackage{slashed}

\allowdisplaybreaks[1]

\newcommand{\T}{{\mathscr T}}
\newcommand{\TT}{{T_{kl,ij}}}

\newcommand{\ac}{{(a)}}
\newcommand{\bc}{{(b)}}

\newcommand{\F}{{\mathcal F}}

\usepackage[normalem]{ulem}  

\begin{document}


\begin{flushright}
\hspace*{60em} 

{\small RBRC-1126\\
RIKEN-QHP-187\\ 
KEK-TH-1814}
\end{flushright}

\title{QCD Kondo effect: Quark matter with heavy-flavor impurities}


\author{Koichi Hattori}
\email[]{koichi.hattori@riken.jp}
\affiliation{RIKEN BNL Research Center, Bldg. 510A, Brookhaven National Laboratory, Upton, NY 11973, USA}
\affiliation{Theoretical Research Division, Nishina Center, RIKEN, Wako, Saitama 351-0198, Japan}

\author{Kazunori Itakura}
\email[]{kazunori.itakura@kek.jp}
\affiliation{KEK Theory Center, Institute of Particle and Nuclear
Studies, High Energy Accelerator Research Organization, 1-1, Oho,
Ibaraki, 305-0801, Japan}
\affiliation{Graduate University for Advanced Studies (SOKENDAI),
1-1 Oho, Tsukuba, Ibaraki 305-0801, Japan}
\author{Sho~Ozaki}
\email[]{sho@post.kek.jp}
\affiliation{KEK Theory Center, Institute of Particle and Nuclear
Studies, High Energy Accelerator Research Organization, 1-1, Oho,
Ibaraki, 305-0801, Japan}
\author{Shigehiro~Yasui}
\email[]{yasuis@th.phys.titech.ac.jp}
\affiliation{Department of Physics, Tokyo Institute of Technology, Tokyo 152-8551, Japan}



\begin{abstract}
We show that the Kondo effect occurs in light quark matter 
which contains heavy quarks
as impurities. 
We consider a scattering between a heavy-flavor impurity and a light quark near a Fermi surface which is mediated by gluon-exchange interactions. 
We find that the scattering amplitude has a logarithmic infrared divergence originating from 
imperfect cancellation between quark-impurity and hole-impurity scatterings in a loop integral, 
implying the presence of a strongly coupled regime near the Fermi surface. Renormalization group method is used to find the Kondo scale 
where a running coupling constant hits a Landau pole. 
Following an illustration by a simple contact-interaction model, 
we examine gluon-exchange interactions on the basis of high density QCD. 
\end{abstract}

\pacs{12.38.Mh,21.65.Qr,12.39.Hg}
\keywords{Quark matter, Kondo effect, Heavy quark effective theory}

\maketitle

\section{Introduction}
\label{sec:introduction}

Infrared instabilities near Fermi surface of degenerated Fermi gas are 
seen in many quantum systems. 
Superconductivity/superfluidity is a well-known example, which appears not only in electron and atom systems but also in nuclear matter and quark matter \cite{Alford:2007xm}.
Also the Kondo effect is caused by an infrared instability near the Fermi surface 
when interactions between degenerated fermions and sufficiently heavy impurities 
have non-Abelian nature (e.g., spin-flip interaction) \cite{1964PThPh..32...37K,1997kphf.book.....H}. 
The Kondo effect induces an enhancement in scattering amplitudes 
between a light fermion and a heavy impurity 
no matter how weak an elementary coupling strength is. 
The Kondo effects are observed in experiments as a change of resistivity 
in a variety of systems from alloys to quantum dots.
To investigate the Kondo effect, a number of theoretical methods has been developed: 
resummation of leading logarithm \cite{Abrikosov1965}, 
the scaling-law 
approach \cite{0022-3719-3-12-008}, 
the numerical renormalization group (RG) method \cite{RevModPhys.47.773}, 
and so on.

The Kondo effect 
occurs when a system is characterized by the following 
four ingredients \cite{Yamada} : 
(0) heavy impurity, (1) Fermi surface, (2) quantum loop effect, and (3) non-Abelian interaction. 
These can be indeed identified in finite density QCD 
with dilute heavy quarks and thus the Kondo effect may take place, 
as recently suggested by one of the present authors \cite{Yasui:2013xr}. In Ref.~\cite{Yasui:2013xr}, 
emergence of the infrared instabilities 
was found in the scattering amplitude between a light fermion and a heavy impurity. 
The scatterings with impurities are governed by strong interaction 
involving several internal degrees of freedom responsible for the non-Abelian properties, such as an isospin for a $\bar{D}$ ($B$) meson in nuclear matter 
and the color for a charm (bottom) quark in quark matter. 
Note that, while a (pseudo-)spin flip plays a role in electron systems to induce a non-Abelian interaction, spin-flip interactions are suppressed in the large quark mass limit in QCD, leaving frozen spin degrees of freedom 
\cite{Isgur:1989vq,Isgur:1989ed,Isgur:1991wq,Manohar:2000dt,Neubert:1993mb}. 
Nevertheless, QCD involves complex non-Abelian interactions associated with 
the isospin and the color degrees of freedom, and they give rise to intriguing phenomena in QCD matter. 
We call the Kondo effect whose non-Abelian property is provided by strong interaction 
the ``QCD Kondo effect.''
When the Kondo effect exists in nuclear and quark systems, 
it will cause a large modification in the transport properties 
and excitations near the Fermi surface. 
Exploring the Kondo effect is thus useful to probe the nuclear and quark matter states.

We investigate the quark matter composed of light (up, down and strange) quarks 
at high density with the heavy (charm or bottom) quarks embedded as impurities. 
We show that a strength of interactions between the light-quark gas and heavy-quark impurity 
grows near the Fermi surface regardless of a weak coupling regime in high density QCD where the asymptotic freedom works. 
To this end, we deal with a free light-quark gas 
and dilute enough heavy-quark impurities, 
so mutual interactions between the heavy quarks are negligible. 
The non-Abelian interaction between the impurity and the scattering fermions, 
which is important in the Kondo effect, is fulfilled by the color $\mathrm{SU}(n)$ symmetry ($n\!=\!3$) 
causing the color-flip in the gluon-exchange interaction. 
The heavy and light quarks belong to the fundamental representation (${\mathbf 3}_{\mathrm{c}}$), 
 and a one-gluon exchange between two quarks $i$, $j$ gives 
a product of color operators $\vec{t}_{i} \cdot \vec{t}_{j}$ 
with $\vec{t}_{k} = (t^{1}_{k},\dots,t^{n^{2}-1}_{k})$ being the generators of 
the color $\mathrm{SU}(n)$ group for the quark $k$. 
This operator gives the non-Abelian interaction between the heavy quark and the light quark.
While the color $\mathrm{SU}(3)$ in quark systems is different from 
the conventional (pseudo-)spin $\mathrm{SU}(2)$ in electron systems, 
we will find that QCD governed by the color $\mathrm{SU}(3)$ 
nevertheless contains all the necessary conditions for the Kondo effect to emerge. 

It is important in the gluon-exchange interaction that the {\it electric} and {\it magnetic} components 
have different properties at finite density.
The electric component gives the short-range interaction 
thanks to the screening effect with the Debye mass in the long-range limit.
The magnetic component, on the other hand, remains a long-range interaction, 
because it is affected only by the dynamical screening \cite{Baym:1990uj}. 
In the present work, we include finite-range gluon exchanges 
with this difference in electric and magnetic components taken into account. 
With this respect as well, we improve the preceding work by contact interactions \cite{Yasui:2013xr}. 
The unscreened magnetic component plays an important role in color superconductivity, 
because it modifies a parametric dependence 
of the ``energy gap" $\Delta$ on the QCD coupling constant $g$: 
the dependence changes from $\Delta \propto {\rm e}^{-c/g^2}$ (the screening case) 
to $\Delta \propto {\rm e}^{-c'/g}$ (the unscreening case) with some constants $c$ and $c'$ \cite{Son:1998uk,Hsu:1999mp} (see also Refs.~\cite{Alford:2007xm,Fukushima:2010bq}). 
Thus, $\Delta$ is estimated to be parametrically large, 
when effects of the magnetic component are taken into account. 
We also investigate effects of the magnetic gluon exchange on the Kondo effect 
in the present work. 

In this work, we apply the RG method, which has been used for the color superconductivity \cite{Evans:1998ek,Son:1998uk,Hsu:1999mp}.
We follow the perturbative calculation by assuming that the interaction coupling is 
small at the energy scale near the Fermi surface.
Considering the scattering amplitude between the heavy quark and the light quark up to one-loop level, we will
investigate how the scattering amplitude depends on the energy scale, and
will estimate the Kondo scale in which the scattering amplitude becomes divergent and the system becomes a  strongly coupled one.

The article is organized as follows.
In Sec.~\ref{sec:Kondo_introduction}, we present a toy-model analysis 
for the Kondo effect by using the RG with a contact interaction. 
In Sec.~\ref{sec:QCD_Kondo_effect}, we describe the formalism of the gluon-exchange interaction 
in high density QCD and of heavy quark effective theory. 
We compute the scattering amplitude up to one-loop level and analyze the RG equation with the gluon exchange interaction. The final section is devoted to the discussions and the conclusion.

\begin{figure}[t!]
 \begin{minipage}{1.0\hsize}
  \begin{center}
    \includegraphics[angle=0,width=70mm]{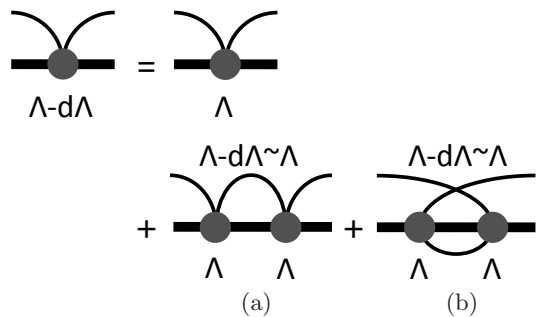}
  \end{center}
 \end{minipage}
\hspace*{2.4cm}(a) \hspace{2.2cm}(b)
\vspace{-0.3cm}
 \caption{Diagrammatic representation of a RG equation in a toy model. Thin and thick lines denote the propagators of light and heavy quarks, respectively. Grey blobs indicate scattering amplitudes at each energy scale.}
\label{fig:Kondo}
\end{figure}

\section{Kondo effect in toy model}
\label{sec:Kondo_introduction}

For a simple illustration of the Kondo effect realized by color-exchange interactions, we briefly discuss a logarithmic infrared divergence 
in the scattering amplitude at one-loop level \cite{1964PThPh..32...37K} 
and then resummation of the leading logarithm 
by a scaling approach \cite{0022-3719-3-12-008}. 
We compute a scattering process between an impurity ($\Psi$) and a scattering light fermion ($\psi$) 
described by a simple contact-interaction Hamiltonian 
\begin{eqnarray}
{\cal H}_{\mathrm{int}} = 
G \sum_{c} \sum_{kl,ij} \psi^{\dag}_{k}(t^{c})_{kl}\psi_{l} \Psi^{\dag}_{i}(t^{c})_{ij}\Psi_{j},
\label{eq:toy_interaction}
\end{eqnarray}
where 
$t^{c}$ ($c=1,\dots,n^2-1$) are the generators of the internal $\mathrm{SU}(n)$ symmetry \cite{Yasui:2013xr}.
We keep a general integer $n$ in the following analysis, but the symmetry group is supposed to be 
the color $\mathrm{SU}(3)$ 
group. Therefore, we will find that the Kondo effect is induced by the color-flipping interaction, 
although spin-flipping process, conventionally discussed in electron systems, 
is suppressed in the large quark mass limit 
\cite{Isgur:1989vq,Isgur:1989ed,Isgur:1991wq,Manohar:2000dt,Neubert:1993mb}. 
The sign of $G$ is chosen so that the interaction vertex at the tree-level gives an attraction for the $\bar{\mathbf{3}}_{\mathrm{c}}$ channel and 
a repulsion for the ${\mathbf{6}}_{\mathrm{c}}$ channel for $n=3$.
When the coupling is small,  the perturbative calculation may be applied 
to the scattering amplitude.
Both the impurity and the light fermion belong to the fundamental representation of the internal $\mathrm{SU}(n)$ symmetry; $k$ and $l$ are the indices 
for the final and initial states of the scattering light fermion, and $i$ and $j$ are 
those of the impurity ($k,l,i,j=1,\dots,n$).

\subsection{Logarithmic infrared divergence}

To investigate an instability near the Fermi surface, 
we focus on scatterings of light fermions on the Fermi surface 
in the initial and final states. 
The scattering amplitude is shown in Fig.~\ref{fig:Kondo} up to one-loop level .
At the tree level, the scattering amplitude is simply given by 
\begin{eqnarray}
 {M}^{(0)}_{kl,ij} = G  \TT
,
 \label{eq:toy_tree}
\end{eqnarray}
with a product of the color matrices 
\begin{eqnarray}
T_{kl,ij} = \sum_c (t^c)_{kl}(t^c)_{ij}
.
\end{eqnarray}
At the one-loop order, the scattering amplitude 
between a light quark and a {\it heavy-quark impurity}
\footnote{We emphasize the four ingredients necessary for the realization of the Kondo effect by italic letters.} 
acquires a {\it quantum loop effect} as (see Appendix A for derivation) 
\begin{eqnarray}
\! {M}^{(1)}_{kl,ij}\!
 &=&\! G^{2} \T^\ac_{kl,ij}\! \int \!\! \cfrac{ \rho(E) }{-E+i\varepsilon}\, {\mathrm{d}}E
 +G^{2} \T^\bc_{kl,ij}\! \int \!\! \cfrac{\rho(E) }{E-i\varepsilon}\, {\mathrm{d}}E ,
 \nonumber
 \\
 \label{eq:toy_scattering_amplitude_1}
\end{eqnarray}
where the first and second terms come from the one-loop diagrams in the second row in Fig.~\ref{fig:Kondo}, respectively, 
and $\T^\ac_{kl,ij}$ and $\T^\bc_{kl,ij}$ are noncommutative products 
of the color matrices on the interaction vertices 
\begin{eqnarray}
\hspace{-0.5cm}
\T^\ac_{kl,ij} &=&
\sum_{c,d} \ \sum_{k'} (t^{c})_{kk'} (t^{d})_{k' l} 
\sum_{i'} (t^{c})_{ii'} (t^{d})_{i'j} 
, 
 \label{eq:identity_10} 
\\
\hspace{-0.5cm}
\T^\bc_{kl,ij} &=& 
\sum_{c,d} 
\ \sum_{k'} (t^{c})_{kk'} (t^{d})_{k' l} 
 \sum_{i'} (t^{d})_{ii'} (t^{c})_{i'j}\, .
\label{eq:identity_20}
\end{eqnarray}
Infinitesimal complex displacements $\pm i \varepsilon$ specify boundary conditions as usual, and 
the first and second terms on the right-hand side 
correspond to the particle and hole propagations, respectively.
$E$ is the absolute value of the energy measured from the Fermi surface ($E \ge 0$). 
A density of states $\rho(E)$ at energy $E$ can be expanded near the Fermi surface 
as $\rho(E) \sim \rho_0 + {\mathcal O}(E)$. The constant term $\rho_0$ is finite in the presence of the {\it Fermi surface}, and thus the numerators of the both terms are finite in the infrared regime $E \sim 0$. Therefore, we find logarithmic infrared divergences as a Fermi surface effect. 
By decomposing the color matrices 
in Eqs.~(\ref{eq:identity_10}) and (\ref{eq:identity_20}) 
with the help of the identities, 
\begin{eqnarray}
\hspace{-0.5cm}
\T^\ac_{kl,ij} 
&=& \cfrac{1}{2} \left( 1-\cfrac{1}{n^2} \right) \delta_{kl}\delta_{ij} 
- \cfrac{1}{n} \TT
, 
 \label{eq:identity_1} 
\\
\hspace{-0.5cm}
\T^\bc_{kl,ij} 
&=& \cfrac{1}{2} \left( 1-\cfrac{1}{n^2} \right) \delta_{kl}\delta_{ij} 
 - \left( \cfrac{1}{n} - \frac{n}{2} \right) 
\TT
,
\label{eq:identity_2}
\end{eqnarray}
we find that the scattering amplitude (\ref{eq:toy_scattering_amplitude_1}) becomes
\begin{eqnarray}
 {M}^{(1)}_{kl,ij}
 &\simeq & G^{2} \rho_0 \left( -\T^\ac_{kl,ij} + \T^\bc_{kl,ij} \right) 
\int \! \cfrac{ {\mathrm{d}}E }{E-i\varepsilon}\nonumber \\
 &=& G^{2} \rho_0 \, \cfrac{n}{2} \, \TT \int \! \cfrac{ {\mathrm{d}}E }{E-i\varepsilon}
\, .
\label{eq:toy_scattering_amplitude_2}
\end{eqnarray}
Here we find that, 
since the decomposed color matrices in Eqs.~(\ref{eq:identity_1}) and (\ref{eq:identity_2}) 
contain the same terms, 
there is a large cancellation between the two terms in 
Eq.~(\ref{eq:toy_scattering_amplitude_1}), i.e., the two one-loop diagrams in Fig.~\ref{fig:Kondo}. 
Only the last term in Eq.~(\ref{eq:identity_2}) survives to give the second line in Eq.~(\ref{eq:toy_scattering_amplitude_2}). 
As mentioned above, this term has a logarithmic infrared divergence. 
If the matrices $\sum_c (t^c)_{kl}(t^c)_{ij}$ 
in the Hamiltonian (\ref{eq:toy_interaction}) were an Abelian type $\delta_{kl}\delta_{ij}$, 
the infrared divergence in the one-loop calculation would completely cancel out 
and there were no infrared divergence. 
Therefore, the infrared divergence surviving in Eq.~(\ref{eq:toy_scattering_amplitude_2}) 
reflects the {\it non-Abelian} properties of the generators $t^{c}$ on the interaction vertices. 
This divergence appears with an arbitrary $SU(n)$ group, with the 
characteristics of the group seen in a simple factor of $n$ in Eq.~(\ref{eq:toy_scattering_amplitude_2}). 

In order to regularize the infrared divergence, 
we introduce an infrared cutoff parameter $\Lambda$.
Then, we can rewrite the scattering amplitude (\ref{eq:toy_scattering_amplitude_2}) as
\begin{eqnarray}
 \hspace{-1em}
 {M}^{(1)}_{kl,ij}(\Lambda) = G^{2} \rho_0 \cfrac{n}{2} \TT 
\int_{E>\Lambda} \cfrac{{\mathrm{d}}E}{E-i\varepsilon}\, . 
\end{eqnarray}
Combining this with the tree-level amplitude (\ref{eq:toy_tree}), 
we obtain the scattering amplitude as 
\begin{eqnarray}
\hspace{-2em}
 {M}_{kl,ij}(\Lambda)
 &=&  {M}^{(0)}_{kl,ij}(\Lambda) + {M}^{(1)}_{kl,ij}(\Lambda) \nonumber \\
 &=& G \TT 
+ G^{2} \rho_0 \cfrac{n}{2} \TT 
\int_{E>\Lambda}  \cfrac{{\mathrm{d}}E}{E-i\varepsilon}\, .
\end{eqnarray}
Kondo showed in Ref.~\cite{1964PThPh..32...37K} 
that the quantum contribution in the second term grows logarithmically 
in the infrared limit $\Lambda \rightarrow 0$, and becomes an important correction to 
the tree-level contribution in the first term. 
This work was extended to relativistic theories 
with the isospin/color symmetry in Ref.~\cite{Yasui:2013xr}. 
While we focused on an illustration of the Kondo effect 
by nonrelativistic expressions in this and the next subsections, 
we will proceed to the relativistic theory on the basis of high-density QCD in the next section.

\subsection{Kondo scale from renormalization group equation}

Resummation of the leading logarithmic contributions can be handled 
by the renormalization group (RG) method. 
Following the description in Ref.~\cite{0022-3719-3-12-008}, 
let us investigate how the scattering amplitude ${M}_{kl,ij}(\Lambda)$ 
behaves as the energy scale $\Lambda$ changes. 
We find a RG equation for ${M}_{kl,ij}(\Lambda)$ as 
\begin{eqnarray}
\hspace{-3em}
&&
\hspace{-0.5cm}
{M}_{kl,ij}(\Lambda-{\mathrm{d}}\Lambda) 
\nonumber \\
&& \hspace{0.2cm}
=  {M}_{kl,ij}(\Lambda)  + G(\Lambda)^{2} \rho_0 \cfrac{n}{2} \TT 
\int^{\Lambda}_{\Lambda-{\mathrm{d}}\Lambda}  \cfrac{{\mathrm{d}}E}{E-i\varepsilon}.
\label{eq:toy_renormalization_group}
\end{eqnarray}
Namely, the scattering amplitude in an energy scale $\Lambda-{\mathrm{d}}\Lambda$ 
is given by a sum of the tree-level contribution at $\Lambda$ and the one-loop 
contribution integrated over a thin shell $\Lambda-{\mathrm{d}}\Lambda \sim \Lambda$. 
Since this quantum correction has the same color-matrix structure as in the tree-level scattering amplitude (\ref{eq:toy_tree}), one can absorb it into a renormalized coupling constant $G(\Lambda) $ 
giving a scattering amplitude in a lower energy scale 
\begin{eqnarray}
{M}_{kl,ij}(\Lambda-{\mathrm{d}}\Lambda) 
 = G(\Lambda-{\mathrm{d}}\Lambda) \TT \, .
\end{eqnarray}
Following this renormalization, 
we obtain a flow equation for the effective coupling constant: 
\begin{eqnarray}
 \Lambda \cfrac{\mathrm{d}}{\mathrm{d}\Lambda} G(\Lambda)
 = -\cfrac{n}{2} G^{2}(\Lambda) \rho_0\, ,
\end{eqnarray}
and its solution 
\begin{eqnarray}
 G(\Lambda) = \cfrac{G(\Lambda_{0})}
{1+\cfrac{n \rho_0}{2} G(\Lambda_{0}) \log \cfrac{\Lambda}{\Lambda_{0}}}
\ .
\end{eqnarray}
Here $\Lambda_{0}$ is an initial energy scale of the RG flow. 
We find that the running coupling constant $ G(\Lambda) $ 
has a Landau pole and diverges 
at $\Lambda = \Lambda_{\mathrm{K}}$, 
where $\Lambda_{\mathrm{K}}$ is the so-called Kondo scale 
\begin{eqnarray}
 \Lambda_{\mathrm{K}} = \Lambda_{0} \exp\left(-\cfrac{2}{n\rho_0 \, G(\Lambda_{0})}\right).
\end{eqnarray}
The existence of the infrared divergence 
indicates that the system is a strongly interacting one near the Fermi surface, 
even though the interaction is weak in the ultraviolet energy scale. 
The Kondo effect arises from four ingredients 
as we mentioned in the Introduction and explicitly showed in this section. Namely, we found logarithmic terms originating from a loop integral in a scattering amplitude between a heavy-flavor impurity and a light quark carrying a Fermi momentum. Although these logarithmic terms cancel out if interactions are Abelian type, 
we showed that the cancellation is not perfect in the presence of non-Abelian interactions. Based on a RG equation, we found a Landau pole of the running coupling constant which becomes divergently large at the Kondo scale near the Fermi surface.

\section{QCD Kondo effect}
\label{sec:QCD_Kondo_effect}


In the last section, we have illustrated the emergence of the Kondo effect 
by using a contact-interaction model. 
We now investigate finite-range gluon interactions on the QCD basis, 
and how an effective four-quark vertex 
shown in Fig.~\ref{fig:Kondo_gluon} 
evolves as the relevant energy scale goes down to the infrared regime near the Fermi surface.

\subsection{Setup of problems}
\label{sec:gluon_exchange_interaction}

\begin{figure}[t,b]
  \begin{center}
\includegraphics[width=0.8\hsize]{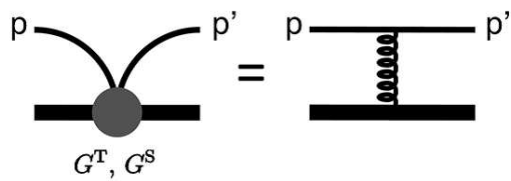}
  \end{center}
\vspace{-0.7cm}
 \caption{
An effective four-quark vertex induced by 
a scattering between a light quark (thin line) and a heavy quark (thick line) 
through a gluon exchange interaction.
}
\label{fig:Kondo_gluon}
\end{figure}

We first look at a gluon propagator at high density. 
While the electric component 
is screened by the Debye mass $m_{\mathrm{D}}$, 
the magnetic component has only the dynamical screening effect 
called the Landau damping \cite{Baym:1990uj}, 
leading to unscreening in an infrared momentum region \cite{Son:1998uk}. 
An explicit form of each component is given by 
\begin{eqnarray}
iD^{\mu\nu}(q)
=
\left\{
\begin{array}{l}
 \cfrac{i}{m_{\mathrm D}^{2}} \hspace{0.5em} {\mathrm{for}} \hspace{0.5em} \mu=\nu=0 
\\
 \cfrac{i (\delta^{ij} -\hat{q}^{i}\hat{q}^{j} )}{ q_{0}^{2} - |\vec{q}\,|^{2} - i \frac{\pi}{2} m_{\mathrm D}^{2} |q_{0}|/|\vec{q}\,| } 
 \hspace{0.5em} {\mathrm{for}} \hspace{0.5em} \mu,\nu=i,j
\end{array}
\right.
\label{eq:gluon_exchange_propagator}
\end{eqnarray} 
with $\hat{q}^{i}=q^{i}/|\vec{q}\,|$, and the off-diagonal components are vanishing. 
Here we have taken the Landau gauge $\xi=0$ otherwise the gauge dependent term would exist $\xi q^\mu q^\nu/q^4$. 
Since we will focus on a small momentum exchange of the order of 
an infrared cutoff scale ($\Lambda \ll k_F$), the momentum dependence 
in the electric component can be ignored in the presence of the Debye mass. 
On the other hand, we need to maintain the momentum dependence in the magnetic component 
since the Landau damping term vanishes as the energy vanishes $q^0 \to 0$.

We suppose that the mass of heavy quarks (charm and bottom) is much larger than the low energy scale in the system.
A useful way to treat the heavy quark is to separate the hard scale momentum 
(of the order of the heavy-quark mass) and the residual momentum $k^\mu$ as 
\begin{eqnarray}
p^{\mu}=m_{\mathrm{Q}}v^{\mu} + k^{\mu},
\end{eqnarray}
with the heavy-quark mass $m_{\mathrm{Q}}$ and the four-velocity $v^{\mu}=(v^0,\vec{v}\,)=(\sqrt{1+|\vec{v}\,|^{2}}, \vec{v}\,)$ 
 \cite{Isgur:1989vq,Isgur:1989ed,Isgur:1991wq,Manohar:2000dt,Neubert:1993mb}.
We note that the spatial component $\vec{v}$ in the four-velocity is different from the normal velocity $\vec{u}$.
Namely, $\vec{v}$ is related to $\vec{u}$ by $\vec{v}=\vec{u}/\sqrt{1-|\vec{u}\,|^2}$.
The residual momentum $k^{\mu}$ is supposed to be much smaller than the heavy-quark mass $k^\mu\ll m_{\rm Q}$ so that the expansion with respect to $1/m_{\rm Q}$ is justified. 
A framework based on the $1/m_{\mathrm{Q}}$ expansion 
is called the heavy quark effective theory (HQET) 
\cite{Isgur:1989vq,Isgur:1989ed,Isgur:1991wq,Manohar:2000dt,Neubert:1993mb},
which has been quite successful in describing heavy quark dynamics in 
a vacuum. 
While $k^\mu$ could be at most of the order of $\Lambda_{\rm QCD}$ in the vacuum, a heavy quark in a medium could typically receive a kick of the order of $m_{\rm D}$ (a typical momentum transfer by one gluon exchange). Therefore, the validity of the HQET in a medium may be given by $ g\mu \ll m_{\rm Q}$.

The leading-order HQET Lagrangian reads 
\begin{eqnarray}
{\cal L}_{\mathrm{HQET}} = 
\bar{Q}_{v} iv \! \cdot \! D Q_{v} + {\cal L}_{\mathrm{light}} + {\cal O}(1/m_{\mathrm{Q}})
.
\end{eqnarray}
A positive-energy component of a quark field is denoted as $Q_v$ which was picked up by a projection 
\begin{eqnarray}
 Q_{v}(x) &=& {\rm e}^{im_{\mathrm{Q}}v \cdot x} \cfrac{1+v\hspace{-0.5em}/}{2} Q(x)
\label{eq:HQ}
,
\end{eqnarray}
where the hard momentum scale is factorized as  
a plane wave, 
and $Q_{v}(x)$ has only a residual momentum $k^{\mu}$. 
While we examine a single heavy flavor below for simplicity, 
a generalization to multiple flavors is straightforward. 
A covariant derivative $D^{\mu}=\partial^{\mu} + ig{\cal A}^{\mu}$ here 
is given by a derivative 
associated with the residual momentum of the heavy quark $k^{\mu}$, 
a coupling constant $g$, and the gluon field ${\cal A}^{\mu}=A^{ c \mu} t^{c}$ 
with $t^{c}$ ($c=1,\dots,8$) being the generators of $\mathrm{SU}(n)$ group 
as in the toy model (\ref{eq:toy_interaction}). 
The QCD Lagrangian for light quarks and gluons is denoted as ${\cal L}_{\mathrm{light}}$. 
When the heavy quark has a finite three-dimensional velocity ($|\vec{v}\,|\neq 0$) in the quark matter, 
the magnetic gluons 
are coupled to the heavy quark dynamics at the leading order, 
because there is a nonvanishing coupling term between 
the spatial components $\vec{v} \cdot \vec{\cal A}$ in $iv \cdot D$, 
while this coupling vanishes as the velocity vanishes ($|\vec{v}\,| = 0$).

As we already emphasized in the discussion on the toy model,
 the heavy-quark spin does not work as an origin of non-Abelian interactions since the spin degrees of freedom are frozen in the heavy quark limit ($m_{\mathrm{Q}} \rightarrow \infty$). 
The Kondo effect is thus induced by the color-flip operators $t^{c}$. 
This may be contrasted with the Kondo effect in condensed matter physics 
where the spin-flip interactions play a crucial role in the electron systems. 

We consider a light quark $q$ scattering off a heavy-quark impurity $Q$ at zero temperature,
\begin{eqnarray}
q^{l}(p)Q^{j} \rightarrow q^{k}(p')Q^{i},
\label{eq:qQ}
\end{eqnarray}
where a momentum exchange $ p^\prime - p$ 
through the finite-range gluon interaction 
gives rise to a momentum dependence of the scattering amplitudes. 
Superscripts of the quark fields denote color indices. 
Let the scattering amplitudes 
be $G^{\rm T}(p^\prime-p)$ and $G^{\rm S}(p^\prime-p)$ as shown in Fig.~\ref{fig:Kondo_gluon}, 
and then effective four-quark vertices are represented by 
\begin{eqnarray}
{\mathcal L}_{\rm eff} &=& 
G^{\rm T}(p^\prime-p) \sum_{a} \bar{q}^k (p') \gamma^{0} v^{0}  (t^{a})^{kl} q^l (p) 
\bar{Q}_{v}^i (t^{a})^{ij} Q_{v}^j 
\nonumber \\
&& 
+ 
 G^{\rm S}(p^\prime-p) \sum_{a} \bar{q}^k (p') \vec{\gamma} \!\cdot\! \vec{v}\,  (t^{a})^{kl} q^l (p) 
\bar{Q}_{v}^i (t^{a})^{ij} Q_{v}^j 
.
\nonumber
\\
\label{eq:scattering_0}
\end{eqnarray}
We have divided the vertices into the temporal (T) and spatial (S) components, because the distinct properties of the electric and magnetic sectors in the gluon propagator (\ref{eq:gluon_exchange_propagator}) distinguish these components. By decomposing $G^{\rm T}$ and $G^{\rm S}$ into the partial waves, we define
\begin{eqnarray}
G^{\rm T}_{\ell} &=& 
\frac{1}{2} \int_{-1}^{1} {\mathrm d}(\cos \theta) \, P_{\ell}(\cos \theta) \, G^{\rm T}(p'-p), \\
G^{\rm S}_{\ell} &=& 
\frac{1}{2} \int_{-1}^{1} {\mathrm d}(\cos \theta) \, P_{\ell}(\cos \theta) \, G^{\rm S}(p'-p),
\end{eqnarray}
where $\theta$ is a scattering angle between the initial and final states of the light quark. 
Since we focus on the S-wave scattering below, 
we suppress the subscripts of 
$G^{\rm T}_{\ell=0}$ and $G^{\rm S}_{\ell=0}$ as $G^{\rm T}$ and $G^{\rm S}$, respectively. 
Then, the S-wave scattering amplitudes are expressed as 
\begin{eqnarray}
\hspace{-4mm}
{\cal M}^{(0)}_{kl,ij}\! =\!
\left( G^{\rm T} v^{0} \gamma^{0} + G^{\rm S} \vec{v} \!\cdot\! \vec{\gamma} \right) 
\sum_{c} (t^{c})_{kl} (t^{c})_{ij} 
\otimes \frac{1 + \slashed v}{2},
\label{eq:scattering_1}
\end{eqnarray}
where spinor indices of the gamma matrices are to be contracted with the light quark fields 
in the initial and final states. The heavy quark fields do not have the four-component spinor structure 
after the projections in Eq.~(\ref{eq:HQ}), 
and thus we do not have to take this structure into account above. 

\subsection{Renormalization group equations} 
\label{sec:RGE_scattering_amplitude}

With the setup in the previous subsection,
we investigate the energy-scale dependence of the scattering amplitudes 
to derive RG equations for the effective four-quark vertices 
originating from the gluon-exchange interactions 
(\ref{eq:gluon_exchange_propagator}) (see Fig.~\ref{fig:Kondo_gluon}). 
The scattering amplitudes are perturbatively
computed in the energy scale near the Fermi surface where 
the QCD coupling constant is sufficiently small at high density.
We will show that the scattering amplitudes have infrared divergences, 
and that 
the RG flows evolve into regimes of strong couplings. 

\begin{figure}[t!]
\vspace{5mm}
\begin{center}
	\includegraphics[width=0.75\hsize]{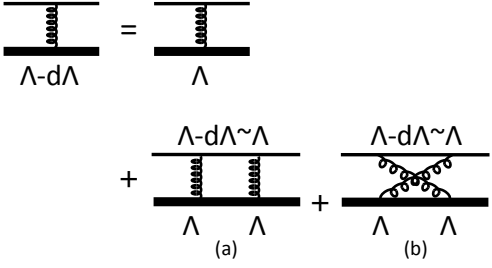}
\end{center}
\vspace{-0.5cm}
 \caption{
Diagrammatic representation of a RG equation with gluon exchange interactions. 
Thin, thick and curly lines denote the propagators of light and heavy quarks, and gluons, respectively. }
\label{fig:Kondo_QCD}
\end{figure}

The RG equation with gluon exchange interactions at one-loop order is depicted in Fig.~\ref{fig:Kondo_QCD}. 
The tree-level scattering amplitude (the first line in Fig.~\ref{fig:Kondo_QCD}) reads 
\begin{widetext}
\begin{eqnarray}
{\cal M}^{(0)}_{kl,ij} &=&
\cfrac{g^2}{2}  \TT  \int_{-1}^{1} {\mathrm d} t \, P_{0}(t) 
\left[ \frac{v^{0}\gamma^{0}}{m_{\mathrm D}^{2}} + \frac{ \frac{2}{3} \vec{v} \!\cdot\! \vec{\gamma}}{(p_{0}-k_{0})^{2} - |\vec{p}-\vec{k}\,|^{2} - i \frac{\pi}{2}m_{\mathrm D}^{2} |p_{0}-k_{0}|/|\vec{p}-\vec{k}\,|} \right] 
\nonumber \\
&=&
g^{2} \TT  \left\{ \frac{1}{m_{\mathrm D}^{2}} v^{0} \gamma^{0} + \frac{1}{2} \frac{2}{3} \left( -\frac{1}{k_{\mathrm F}^{2}} \frac{1}{3} \log \frac{8+i \lambda}{i\lambda}  \right) \vec{v} \!\cdot\! \vec{\gamma} \right\} 
,
\label{eq:scattering_amplitude_0}
\end{eqnarray}
\end{widetext}
where the kinematics is specified by 
\begin{eqnarray}
&& |p'_{0}-p_{0}| = \Lambda, \\
 &&|\vec{p}\,'| = |\vec{p}\,| = k_{\mathrm F}, \\
 &&\hat{p}' \cdot \hat{p} = \cos \theta =t
 ,
\end{eqnarray}
with $\hat{p}\,'=\vec{p}\,'/|\vec{p}\,'|$ and $\hat{p}=\vec{p}/|\vec{p}\,|$. 
Here $\Lambda$ is a small energy scale of the quantum fluctuation, 
as we are interested in an evolution of the RG equation near the Fermi surface. 
The three-dimensional momenta $\vec{p}\,'$ and $\vec{p}$ 
are approximated to be the Fermi momentum $k_{\mathrm F}$.
In the second line in Eq.~(\ref{eq:scattering_amplitude_0}), 
we have defined $\lambda = (\pi/2) m_{\mathrm D}^{2}\Lambda/k_{\mathrm F}^{3}$, 
which is a small quantity ($\lambda \ll 1$) when 
$m_{\mathrm D} \simeq g\mu$ and $\mu \simeq k_{\mathrm F}$.
We find that even the tree-level scattering amplitude ${\cal M}^{(0)}_{kl,ij}$ 
depends on the cutoff parameter $\Lambda$, 
because of the energy-scale dependence $\lambda$ in the magnetic gluon propagator.
The cutoff-parameter $\Lambda$ is an energy measured from the Fermi surface.
We assume that $\Lambda$ is sufficiently smaller than $k_{\mathrm F}$, 
and hence that ${\cal O}((\Lambda/k_{\mathrm F})^{2})$ is neglected 
compared to ${\cal O}(\Lambda/k_{\mathrm F})$.
We can identify tree level contributions to $G^{\rm T}$ and $G^{\rm S}$ from Eq.~(\ref{eq:scattering_amplitude_0}). Using expansion 
with respect to $\lambda$ (Appendix~\ref{sec:integral_formula}), we find 
\begin{eqnarray}
G^{\rm T}_{\mathrm{tree}}
&=&
\frac{g^2}{m_{\mathrm D}^{2}},
\label{eq:A0_tree} \\
G^{\rm S}_{\mathrm{tree}}
&=&
-\frac{g^{2}}{k_{\mathrm F}^{2}} \frac{2}{3}\frac{1}{6} \log \frac{8+i \lambda}{i \lambda}.
\label{eq:B0_tree}
\end{eqnarray}
We again note that the unscreened magnetic gluon induces 
the $\Lambda$ dependence of $G^{\rm S}$ at the tree level, 
which was important in the color superconductivity \cite{Son:1998uk,Hsu:1999mp}.

Next, we proceed to computation of the quantum loop effect at the one-loop level: 
${\cal M}^{(1)}_{kl,ij} = {\cal M}^{(1a)}_{kl,ij} + {\cal M}^{(1b)}_{kl,ij}$,
where ${\cal M}^{(1a)}_{kl,ij}$ and ${\cal M}^{(1b)}_{kl,ij}$ are the contributions from the first and second loop diagrams in Fig.~\ref{fig:Kondo_QCD}.
The scattering amplitude ${\cal M}^{(1a)}_{kl,ij}$ is given by
\begin{eqnarray}
{\cal M}^{(1a)}_{kl,ij}
&=&
i (ig)^{4} \T^\ac_{kl,ij} \int \frac{{\mathrm d}^{4}q}{(2\pi)^{4}}
iS(q) \gamma^{\rho}   v^{\nu} 
\otimes \frac{i(1+v\hspace{-0.5em}/)}{2v \!\cdot\! (p-q)} v^{\sigma} 
\nonumber \\ 
&& \hspace{2.3cm} \times
 iD_{\mu\nu}(q-p)  iD_{\rho\sigma}(q-k)\, . 
\end{eqnarray}
We have the product of the color matrices $\T^\ac_{kl,ij}$ as in Eq.~(\ref{eq:toy_scattering_amplitude_1}), 
which will again play a role below as in the toy model calculation. 
The light-quark propagator at finite density and zero temperature is given by
\begin{eqnarray}
 iS(q) = q \hspace{-0.5em/}
\left[ \frac{i}{q^{2}+i\varepsilon} - 2\pi \delta(q^{2}) \theta(q_{0}) \theta(k_{\mathrm F} - |\vec{q}\,|) \right],
\label{eq:light_quark_propagator}
\end{eqnarray}
where the second term is the Pauli blocking effect \cite{Kaiser2002255}. 
Note that the propagator of the heavy quark, ${i(1+v\hspace{-0.5em}/)}/{2v\!\cdot\! (p-q)}$, 
does not have the Pauli-blocking term, because the heavy quarks are treated as dilute impurities. 
By applying the partial-wave decomposition to the gluon propagators 
$D_{\mu\nu}(q-p)$ and $D_{\mu\nu}(q-k)$, we obtain 
the S-wave part of the scattering amplitude ${\cal M}^{(1a)}_{kl,ij}$ as 
\begin{eqnarray}
{\cal M}^{(1a)}_{kl,ij}
&=&  \T^\ac_{kl,ij} \!
\int_{q \ge k_{\mathrm F}} \!\!\!\!\!\!\!\!\! {\mathrm d}q \, q^{2} \!\!
 \left[  \F_{12} (q) \gamma^{0}
   + \F_{21}(q)\frac{\vec{v} \!\cdot\! \vec{\gamma}}{|\vec{v}\,|}
 \right] 
 \otimes \frac{1+v\hspace{-0.5em}/}{2}  
,
\nonumber 
\\
\label{eq:one_loop_A1}
\end{eqnarray}
where the integrand is given by 
\begin{eqnarray}
\label{eq:F12}
\F_{12} (q) &=&\frac{1}{2(2\pi)^{2}} \bigg\{
v_{0}^{2} F_{1}(q) \left(G^{\rm T}\right)^{2} 
\\
&& \hspace{0.5cm}+ 2v_{0} |\vec{v}\,| F_{2}(q) G^{\rm T}G^{\rm S}  
+ |\vec{v}\,|^2 F_{1}(q) \left(G^{\rm S}\right)^{2} 
\biggr\}\, ,
\nonumber 
\end{eqnarray}
and $\F_{21}(q)$ with an interchange of the subscripts 
($1 \leftrightarrow 2$) in Eq.~(\ref{eq:F12}). 
The coefficient functions we compute below are 
\begin{eqnarray}
\hspace{-2em}
F_{1}(q) &=& \int_{-1}^{1} {\mathrm d}t \frac{1}{v_{0}(p_{0}-q_{0})-\vec{v}\cdot\vec{p}+|\vec{v}\,| |\vec{p}\,| t + i\varepsilon}, 
\label{eq:function_F1} \\
\hspace{-2em}
F_{2}(q) &=& \int_{-1}^{1} {\mathrm d}t \frac{t}{v_{0}(p_{0}-q_{0})-\vec{v}\cdot\vec{p}+|\vec{v}\,| |\vec{p}\,| t + i\varepsilon}. 
\label{eq:function_F2}
\end{eqnarray}
In a similar way, we obtain the S-wave part of the scattering amplitude ${\cal M}^{(1b)}_{kl,ij}$ 
(the second one-loop diagram in Fig.~\ref{fig:Kondo_QCD}) as 
\begin{eqnarray}
{\cal M}^{(1b)}_{kl,ij}
&=&  \T^\bc_{kl,ij} \!
\int_{q \le k_{\mathrm F}} \!\!\!\!\!\!\!\!\! {\mathrm d}q \, q^{2} \!\!
 \left[ \F_{12} (q) \gamma^{0}
   + \F_{21}(q)\frac{\vec{v} \!\cdot\! \vec{\gamma}}{|\vec{v}\,|}
 \right] 
 \otimes \frac{1+v\hspace{-0.5em}/}{2}  
,
\nonumber
\\
\label{eq:one_loop_A2}
\end{eqnarray}
We note that the momentum integral in Eq.~(\ref{eq:one_loop_A1}) is performed 
over the particle state above the Fermi surface due to 
the Pauli blocking effect in Eq.~(\ref{eq:light_quark_propagator}), 
while the momentum integral in Eq.~(\ref{eq:one_loop_A2}) is performed 
over the hole state below the Fermi surface. 
Since these integrals are divergent in the infrared energy region, 
we introduce infrared cutoff parameters $\Lambda>0$, and perform the integrals in Eqs.~(\ref{eq:one_loop_A1}) and (\ref{eq:one_loop_A2}) 
up to $q \ge k_{\mathrm F}+\Lambda$ and $q \le k_{\mathrm F}-\Lambda$, respectively. 
Following from the integration, we obtain the energy-scale dependence of the scattering amplitude as 
\begin{eqnarray}
{\cal M}^{(1)}_{kl,ij}(\Lambda) = {\cal M}^{(1a)}_{kl,ij}(\Lambda) + {\cal M}^{(1b)}_{kl,ij}(\Lambda)
\label{eq:M1}
.
\end{eqnarray}

Now that we have obtained the energy-scale dependence of the scattering amplitude, 
we can derive the RG equations as shown 
by using the toy model in Sec.~\ref{sec:Kondo_introduction} 
[see descriptions below Eq.~(\ref{eq:toy_renormalization_group})]. 
Since the scattering amplitudes (\ref{eq:scattering_amplitude_0}) and (\ref{eq:M1}) 
have the spinor indices as well as the color indices, 
we compare the terms proportional to $\gamma^{0}$ 
and $\vec{v} \cdot \vec{\gamma}$ separately, 
and then obtain the RG equations 
for $G^{\rm T}(\Lambda)$ and $G^{\rm S}(\Lambda)$, respectively as 
\begin{eqnarray}
-v^{0} \frac{{\mathrm d}G^{\rm T}}{{\mathrm d}\Lambda} \TT 
&=&
\T^\ac_{kl,ij} (k_{\mathrm F}+\Lambda)^{2} \F_{12} (k_{\mathrm F}+\Lambda) 
\nonumber
\\
&&
+ \T^\bc_{kl,ij}  (k_{\mathrm F}-\Lambda)^{2} \F_{12} (k_{\mathrm F}-\Lambda) 
\label{eq:renormalization_group_eq1} 
,
\\
- \vert\vec{v}\vert \frac{{\mathrm d}G^{\rm S}}{{\mathrm d}\Lambda} \TT
&=&
\T^\ac_{kl,ij} (k_{\mathrm F}+\Lambda)^{2} \F_{21} (k_{\mathrm F}+\Lambda) 
\nonumber
\\
&&
+ \T^\bc_{kl,ij}  (k_{\mathrm F}-\Lambda)^{2}  \F_{21} (k_{\mathrm F}-\Lambda) 
\nonumber
\\
&&
+ g^{2} \frac{2}{3} \frac{\vert\vec{v}\vert}{6k_{\mathrm F}^{2}} \frac{1}{\Lambda} \TT
\label{eq:renormalization_group_eq2}
,
\end{eqnarray}
where $\F_{12}(q)$ and $\F_{21}(q)$ are evaluated at $q= k_{\mathrm F} \pm \Lambda$. 
The last term in Eq.~(\ref{eq:renormalization_group_eq2}) proportional to $g^{2}$ 
comes from the unscreened magnetic gluon exchange at the tree level.
The last term in r.h.s. in Eq.~(\ref{eq:renormalization_group_eq2}) was derived from the cutoff dependence in the tree level amplitude (\ref{eq:B0_tree}). Here we used the relation $\log(8+i\lambda)/i\lambda \simeq \log 8/\lambda + i(\pi/2+n\pi)$ for small $\lambda \ll 1$ with $n$ being an integer, and left only the real part as the leading term in the infrared limit ($\Lambda \rightarrow 0$).

\subsection{Solution}
\label{sec:solution}

In order to obtain analytic solutions of 
the RG equations (\ref{eq:renormalization_group_eq1}) and (\ref{eq:renormalization_group_eq2}),
we shall examine the coefficient functions $F_{1}(q)$ and $F_{2}(q)$ defined 
in Eqs.~(\ref{eq:function_F1}) and (\ref{eq:function_F2}), respectively. 
The kinematical variables are specified as 
$p_{0} = |\vec{p}\,| = k_{\mathrm F}$, $|\vec{v}\,| = v$, 
$\vec{v} \cdot \vec{p} = vk_{\mathrm F} \cos \alpha$, and $q_{0} = |\vec{q}\,| = q$. 
Then, we expand $F_{1}(q)$ and $F_{2}(q)$ with respect to 
the nonrelativistic velocity of heavy impurities $v \ll 1$, as shown in Appendix~\ref{sec:v}. 
Inserting Eqs.~(\ref{eq:F1+})--(\ref{eq:F2-}) obtained up to ${\mathcal O}(v)$ 
into the RG equations (\ref{eq:renormalization_group_eq1}) 
and (\ref{eq:renormalization_group_eq2}), 
we find corresponding equations at ${\cal O}(v)$ accuracy as 
\begin{eqnarray}
 -\frac{{\mathrm d}G^{\rm T}}{{\mathrm d}x}  
&=&
\frac{k_{\mathrm F}^{2}}{2(2\pi)^{2}} \left[ \, \frac{n}{x}   
+ \frac{\cos\alpha}{x^2} \left( {\mathcal N} - 2 x \right) v
\, \right]  \left(G^{\rm T}\right)^2 
\nonumber
\\
&&+  {\cal O}(v^{2})
\label{eq:dA}
,
\end{eqnarray}
and 
\begin{eqnarray}
 - \frac{{\mathrm d}G^{\rm S}}{{\mathrm d}x}  
&=&
 \frac{k_{\mathrm F}^{2}}{2(2\pi)^{2}} \left[ \, 
\frac{2n}{x} G^{\rm T}G^{\rm S} - \frac{1}{3x^2} \left( {\mathcal N} - n x \right) \left(G^{\rm T}\right)^2
\, \right]
\nonumber 
\\
&& +  g^2 \frac{2}{3} \frac{C}{6k_{\mathrm F}^2} \frac{1}{x}
+ {\cal O}(v^{1})
\label{eq:dB}
,
\end{eqnarray}
where $x=\Lambda/k_{\mathrm F}$ is a normalized cutoff parameter, 
and a coefficient ${\mathcal N} $ is defined by 
\begin{eqnarray}
{\mathcal N}  = \frac{4}{C} \left( 1 - \frac{1}{n^2} \right) +  \left(n-\frac{4}{n}\right) 
,
\end{eqnarray}
with $C$ being an eigenvalue of the Casimir operator $\sum_c (t^{c})_{kl}(t^{c})_{ij}$. 
In case of the color $\mathrm{SU}(3)$ symmetry with $n\!=\!3$, 
for example, the $\bar{\mathbf 3}_{\mathrm{c}}$ channel gives $C=-2/3$, 
and the ${\mathbf 6}_{\mathrm{c}}$ channel gives $C=1/3$. 
When obtaining Eqs.~(\ref{eq:dA}) and (\ref{eq:dB}) 
from Eqs.~(\ref{eq:renormalization_group_eq1}) and (\ref{eq:renormalization_group_eq2}), respectively, 
most of the terms cancel between the contributions from the two diagrams at the one-loop order 
as shown by the toy model [see Eq.~(\ref{eq:toy_scattering_amplitude_2})] 
because of the common structures in the decomposed color matrices 
in Eqs.~(\ref{eq:identity_1}) and  (\ref{eq:identity_2}).

By solving the coupled differential equations (\ref{eq:dA}) and (\ref{eq:dB}), 
we obtain a solution for $G^{\rm T}(x)$ as
\begin{eqnarray}
G^{\rm T}(x) &=& G^{\rm T}(x_0) \Phi (x)
\nonumber
\\
&&+ v \left\{G^{\rm T}(x_0)\Phi (x)  \right\}^2
\left(  \zeta (x)+ \cfrac{k_{\mathrm F}^{2}}{(2\pi)^2}  \log \cfrac{x}{x_0} \right) \cos \alpha
\nonumber \\
&& +
{\cal O}(v^2),
\label{eq:solution_A}
\end{eqnarray}
where an initial value $G^{\rm T}(x_0)$ of the flow is given at 
a scale $x_0 = \Lambda_{0}/k_{\mathrm F}$, 
and we introduced 
\begin{eqnarray}
&&
\Phi(x)=
\cfrac{1}{1+G^{\rm T}(x_0) \cfrac{n}{2} \cfrac{k_{\mathrm F}^2}{(2\pi)^2} \log \cfrac{x}{x_0} } 
\label{eq:Phi}
\, ,
\\
&&
\zeta(x) = \cfrac{k_{\mathrm F}^{2}}{2(2\pi)^2} 
{\mathcal N} \left( \frac{1}{x} - \frac{1}{x_0} \right) 
.
\nonumber
\\
\end{eqnarray}
We also obtain a solution for $G^{\rm S}(x)$ as 
\begin{eqnarray}
G^{\rm S}(x) &=& - \frac{n}{2} {\mathcal G} 
- \frac{n}{2} \left( G^{\rm T}(x_0) + 
{\mathcal G} \right) \Phi (x)
\nonumber \\
&&
+ \frac{n}{2} \biggl\{ \Big( 1 + 
\zeta (x) \Big) G^{\rm T}(x_0) 
+ 3G^{\rm S}(x_0) - 
{\mathcal G}  \biggr\} \Phi^2(x)
\nonumber \\
&& 
+ {\cal O}(v)
,
\label{eq:solution_B}
\end{eqnarray}
where the terms originating from the Landau damping are proportional to 
\begin{eqnarray}
{\mathcal G} = \frac{2}{3} \frac{g^2}{6k_{\mathrm F}^2} \log\frac{x}{x_0}
\ .
\end{eqnarray}
Note that, since a factor of $v$ was extracted in Eq.~(\ref{eq:scattering_1}), 
$G^{\rm S}(x)$ is ${\cal O}(1)$ when the scattering amplitude is ${\cal O}(v)$ 
and is absent when the heavy impurities are at rest ($v = 0$). 

Interestingly, we find that $\Phi$ defined in Eq.~(\ref{eq:Phi}) 
grows logarithmically as the infrared cutoff $\Lambda$ approaches the Fermi surface. 
Therefore, both of the effective four-Fermi interactions $G^{\rm T}(x)$ and $G^{\rm S}(x)$ 
diverge at the Kondo scale 
\begin{eqnarray}
\Lambda_{\mathrm{K}} = \Lambda_0 \exp \left( - \cfrac{2(2\pi)^2}{nk_{\mathrm F}^2 G^{\rm T}(x_0)} \right),
\end{eqnarray}
which is smaller than an initial value $\Lambda_{0}$.
When the scattering amplitude at the initial energy scale, say $\Lambda_{0} \simeq k_{\mathrm{F}}$ ($x_{0} \simeq 1$), is given by $G^{\rm T}(x_{0}=1)\simeq g^2/k_{\mathrm F}^2$, 
this divergence indeed emerges in the infrared regime 
since the Kondo scale is found to be much smaller than the Fermi momentum as 
\begin{eqnarray}
\Lambda_{\mathrm{K}} \simeq k_{\mathrm F} \exp \left( - \cfrac{2(2\pi)^2}{ng^2} \right) 
\ll k_{\mathrm F}.
\end{eqnarray}
We note that the Kondo scale is common to several channels of $\mathrm{SU}(n)$ symmetry, such as $\bar{\mathbf{3}}_{\mathrm{c}}$ and $\mathbf{6}_{\mathrm{c}}$ for $n=3$. 
It is also straightforward to extend the above RG analysis to a light-quark matter 
which contains heavy-antiquark impurities, and we find the same Kondo scale.

The existence of the divergence in the infrared region indicates that 
the strengths of the four-quark vertices grow logarithmically, and 
the system goes into 
a strongly coupled one at low energy scale. 
The naive perturbation breaks down in such a regime 
as inferred by the emergence of the Kondo scale or the Landau pole. 
When $n=3$, for example, the attractive and repulsive nature 
in $\bar{\mathbf{3}}_{\mathrm{c}}$ and $\mathbf{6}_{\mathrm{c}}$ channels, respectively, 
are strongly enhanced in the infrared regime. 
This result is not affected by the existence of the long-range property in the magnetic gluon.
It is also important to note that the infrared divergence appears 
not only in ${\cal O}(v^{0})$ but also in ${\cal O}(v^{1})$ 
in the scattering amplitudes $G^{\rm T}$ and $G^{\rm S}$ 
as shown in Eqs.~(\ref{eq:solution_A}) and (\ref{eq:solution_B}).


Since effects of temperature smear the Fermi surface, 
we could regard the infrared cutoff energy near the Fermi surface as a temperature. 
In this sense, the physical interpretation of the Kondo scale $\Lambda_{\mathrm{K}}$ 
is regarded as the Kondo temperature at which the scattering amplitudes become divergent.
This observation is consistent with the original analysis performed by Kondo in Ref.~\cite{1964PThPh..32...37K}.

\section{Summary and discussion}
\label{sec:summary}



In summary, we have shown that the Kondo effect is realized in quark matter with heavy quark impurities through the RG analysis of the scattering amplitudes between a light quark and a heavy-flavor impurity. 
The emergence of the Kondo effect in quark matter is natural because this system possesses four necessary ingredients for the Kondo effects to occur, namely, heavy impurities, Fermi surface, quantum loop effect, and non-Abelian interactions. We have analyzed the scattering amplitudes with color-exchange interactions which are given by the (screened) electric and the (unscreened) magnetic gluon exchanges. By solving the RG equations for the effective interaction strengths $G^{\rm T}$ and $G^{\rm S}$, 
we found that there appears the Kondo scale at which the scattering amplitude diverges, and that the system becomes a strongly interacting one. 
We also found that the unscreened magnetic component in the gluon exchange 
does not contribute to the Kondo scale, 
but that it is important when a heavy impurity is moving at a nonzero velocity, which may be a more realistic situation in quark matter created in low-energy heavy-ion collisions.

The Kondo effect is clearly a new idea in the field of QCD and will provide us with a novel viewpoint on the physics of high density quark matter. We conclude the paper by listing some of the interesting problems which we can investigate with this new picture. 
\\

\vspace{-2mm}

\noindent
{\bf (I)} {\it Transport properties of quark matter at low temperature}\\
Increasing strength of the scattering amplitude implies a change of mobility of light quarks and thus will bring about significant change in the transport properties of quark matter. 
For example, similar to the ordinary Kondo effect in condensed matter, 
the QCD Kondo effect will increase the resistivity of light quarks in a quark-gluon plasma at low temperature. Also, we can expect that transport coefficients such as shear viscosity could be affected by the QCD Kondo effect. 
Furthermore, since the effect will be enhanced in proportion to the number density of charm and/or bottom quarks, we expect that anisotropic behavior in quark matter could be generated by inhomogeneous distribution of the heavy impurities. We will be able to study these problems experimentally 
in low-energy heavy-ion collisions at RHIC, GSI-FAIR, and J-PARC 
which will create high density quark matter with charm quark impurities at moderate temperature. 
We also propose to study the Kondo effect and the modification of transport properties in lattice simulation for two-color QCD, which is free from the sign problem. Recall that the Kondo effect should occur for non-Abelian SU($n$) groups including the two-color QCD with $n=2$.
\\

\vspace{-2mm}

\noindent
{\bf (II)} {\it Kondo effect induced by heavy antiparticles}\\
It is straightforward to consider the Kondo effect when the heavy impurity is replaced by a heavy antiquark. 
We expect, however, that heavy quarks and heavy antiquarks play different roles in the Kondo effect because the matter is made of light quarks and the interaction between light quarks and heavy antiquarks should be different (we will discuss this point also in the next topic). In heavy-ion collisions, heavy quarks will be created together with heavy antiquarks. However, if heavy quarks and heavy antiquarks show different Kondo effects, we should be able to detect the effect of finite density or the existence of a quark Fermi surface. 
\\


\vspace{-2mm}

\noindent
{\bf (III)} {\it Possible formation of the Kondo-Yosida state}\\
The strongly enhanced scattering amplitudes in the Kondo effect may suggest even the formation of a bound state of the light quark and the heavy quark. 
Such a bound state is discussed in the ordinary Kondo effect and 
is called the Kondo-Yosida (KY) state \cite{PhysRev.147.223}. 
This is a bound state of the impurity and the light particle-hole pairs around the Fermi surface, which is induced by the enhanced interaction in an attractive channel (e.g., the spin-singlet channel in spin-spin interaction) in the Kondo effect.
In quark matter, the heavy quark $Q$ 
 may lead to the formation of the color antitriplet ($\bar{\mathbf{3}}_{\mathrm{c}}$) KY state, given by the heavy quark and the light quark-hole pairs.
In a similar way, the heavy antiquark $\bar{Q}$ 
 may induce the formation of the color singlet (${\mathbf{1}}_{\mathrm{c}}$) KY state.
It may be interesting to note that the $\mathbf{3}_{\mathrm{c}}$ KY state is colorful, while the $\mathbf{1}_{\mathrm{c}}$ KY state is colorless.
This difference may affect the hadronization properties of the heavy quark and antiquarks 
in a quark matter produced in the relativistic heavy ion collisions.\\

\vspace{-2mm}

\begin{table} 
\caption{The Kondo effect in quark matter and nuclear matter for a heavy quark ($Q$) and a heavy antiquark ($\bar{Q}$).
 The channels in which the Kondo effect occurs are checked by $\surd$.}
\begin{ruledtabular}
\begin{tabular}{ccc}
 & quark matter & nuclear matter   \\
\hline
 heavy-quark & $Q$ $\surd$ & $\Lambda_{\mathrm{Q}}$  \\
 heavy-antiquark & $\bar{Q}$ $\surd$ & $P_{\bar{\mathrm{Q}}}$, $P_{\bar{\mathrm{Q}}}^{\ast}$ $\surd$  \\
\end{tabular}
\end{ruledtabular}
 \label{table:nuclear_quark}
\end{table}

\noindent 
{\bf (IV)} {\it Kondo effect in hadronic matter}\\
It is an important question to ask whether the Kondo effect can occur in hadronic matter (Table~\ref{table:nuclear_quark}).
For the heavy quark $Q$, the heavy baryon $\Lambda_{Q}$, which is the lowest state $qqQ$ baryon with spin-parity $1/2^+$ and isospin singlet ($qq=ud$), is 
the effective degree of freedom in nuclear matter\footnote{We should recall that $\bar{q}Q$ meson in nuclear matter is an excited state, and is not considered in the ground state.}.
Then, the non-Abelian interaction for $\Lambda_{Q}$ can be given by isospin- and spin-dependent interactions. 
However, there is indeed no such interaction, because $\Lambda_{\mathrm{Q}}$ carries no isospin and the spin-flip interaction is suppressed by $1/m_{\mathrm{Q}}$ for large mass of heavy quark $m_{\mathrm{Q}}$. 
Thus, we expect there will be no Kondo effect for the $\Lambda_{\mathrm{Q}}$ baryon in nuclear matter.

On the other hand, for the heavy antiquark $\bar{Q}$, the heavy mesons $P_{\bar{\mathrm{Q}}}$ and $P_{\bar{\mathrm{Q}}}^{\ast}$, where $P_{\bar{\mathrm{Q}}}$ ($P_{\bar{\mathrm{Q}}}^{\ast}$) is the ground state of a heavy meson $q\bar{Q}$ with spin-parity $0^-$ ($1^{-}$) and isospin doublet, exist as the effective degrees of freedom in nuclear matter
\footnote{
We note that $P_{\bar{\mathrm{Q}}}$ and $P_{\bar{\mathrm{Q}}}^{\ast}$ are degenerate in mass in the heavy quark limit ($m_{\bar{\mathrm{Q}}} \rightarrow \infty$), and hence both of them should be considered.}.
Interestingly, the interaction between a $P_{\bar{\mathrm{Q}}}$ ($P_{\bar{\mathrm{Q}}}^{\ast}$) meson and a nucleon $N$ has the isospin exchange induced by the isospin symmetry $\mathrm{SU}(2)$, and it can cause the Kondo effect. 
This was indeed demonstrated by using the contact interaction in the previous work \cite{Yasui:2013xr}.\\

\vspace{-2mm}

\noindent
{\bf (V)} {\it Possible continuity of Kondo-Yosida state from quark matter to hadronic matter}\\
In the previous two subsections, we have discussed that the heavy-quark ($Q$) can lead to the KY state in the quark matter, and not in the nuclear matter.
On the other hand, the heavy-antiquark ($\bar{Q}$) can lead to the KY state in both phases.
We expect that the KY state for $Q$ can have a continuous change, when the phase changes from the quark matter to the hadronic matter.
This is analogous to the hadron-quark continuity discussed in the color-flavor locked (CFL) phase in color superconductivity \cite{Alford:2007xm}.\\

\vspace{-2mm}

\noindent
{\bf (VI)} {\it Competition between the Kondo effect and the color superconductivity}\\
It is interesting to consider the relationship between the Kondo effect and the color superconductivity.
As analyzed in the present study, the Kondo effect is given by the existence of the ungapped Fermi surface.
When the color superconductivity is realized, the Cooper pairs ($qq$) leads to the disappearance of the Fermi surface due to the energy gap, and hence the Kondo effect becomes suppressed.
Nevertheless, when the $qQ$ interaction is much stronger than the $qq$ interaction, the Kondo effects may overcome the color superconductivity.
In the CFL phase, all the quark pairs ($ud$, $ds$, $su$) acquire the energy gap in three flavor case, and hence the Kondo effect will be competitive to CFL phase.
In the two flavor case (2SC), however, only the $ud$ pairs acquire the gap and $ds$ and $su$ pairs are not suffered from the Cooper pair formation,
and hence the Kondo effect will be still realized for those free quarks.
In any case, the competition between the Kondo effect and the color superconductivity will be an important subject in future study.\\

\vspace{-2mm}

\noindent
{\bf (VII)} {\it Gauge invariance of the QCD Kondo effect}\\
All the calculation in Sec.~\ref{sec:QCD_Kondo_effect} was done in the Landau gauge (ƒÌ = 0). 
Then there comes a natural question how the gauge dependence appears in the results, e.g., the Kondo scale. 
Let us first notice that the essence of the QCD Kondo effect 
can be captured by using the four-Fermi interaction as we showed in Sec.~\ref{sec:Kondo_introduction}, 
implying that the gauge dependence in QCD calculation, if any, is expected to be 
a subdominant effect on the Kondo scale. 
In Sec.~\ref{sec:QCD_Kondo_effect} for the QCD calculation, one 
can explicitly show that at least the tree level contribution to the $qQ$ scattering amplitude Eq.~(\ref{eq:scattering_amplitude_0}) does not depend on the gauge parameter $\xi$ (which is easily verified with the use of equations of motion for light quarks: the gauge dependent vertex factor $\bar u(p_2)g\gamma^{\mu}\xi q_{\mu}q_\nu u(p_1)$ with $q^\mu = p_2^\mu -p_1^\mu$ is vanishing due to $(\slash {\! \! \! p}-m)u(p)=0$). The same argument does not hold for one-loop diagrams and thus it seems that there remains gauge-parameter dependence. Nevertheless we expect that the Kondo scale may be gauge invariant from the experience of the gap in color superconductivity. While the gap in color superconductivity apparently depends on the gauge parameter $\xi$ at higher orders, it is proven to be gauge invariant \cite{PR,GR,HWR}. 
\\

\vspace{-2mm}

\noindent
{\bf (VIII)} {\it Effects of self-energy}\\
In the present paper, we focused only on the effects induced by the presence of heavy quarks 
embedded in a light quark matter as impurities. 
However, it is pointed out that the quark matter itself becomes ``non-Fermi liquid" 
due to logarithmic enhancement of the self-energy of light quarks 
within the framework of hard-dense loop approximation \cite{Schaefer,Schaefer2}. 
Such a logarithmic enhancement may lead to a breakdown of perturbation theory 
at very low energies whose parametric dependence on the coupling constant 
is similar to the Kondo scale: $\exp(-a/g^2)$. 
For the light-heavy quark scattering amplitude discussed in the present paper, 
the diagrams 
with insertions of the fermion's self-energies appear at higher orders 
in the naive power counting. 
Although the origin of the logarithmic enhancement is different from that of the Kondo effect, 
one may need to include both effects for a complete description of the scattering amplitude, 
which will additionally require a wave-function renormalization for a light quark 
(logarithmic enhancement will be absent for the heavy quark field 
included as a dilute impurity, 
since it does not form a Fermi surface 
and has a large mass scale regularizing the infrared dynamics). 
It is interesting enough to investigate how the Kondo effect 
is modified by the inclusion of non-Fermi liquid effects of a light quark matter.

\begin{acknowledgments}
This work is supported in part by the Grants-in-Aid for 
Scientific Research from JSPS (Grant No. 25247036) (S.~Y.). 
The research of K.H. is supported by JSPS Grants-in-Aid No.~25287066. 
\end{acknowledgments}

\vspace{0.8cm}

\appendix 

\section{One loop contribution in toy model} 
\label{sec:loop_M}

\begin{figure}[t!]
	\begin{center}
		\includegraphics[width=0.8\hsize]{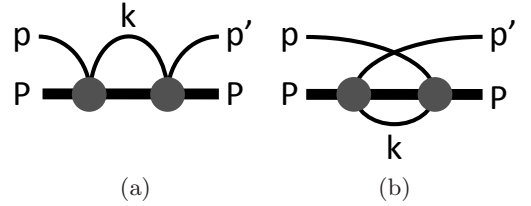}
	\end{center}
\vspace{-0.4cm}
(a) \hspace{2.8cm} (b)
 \caption{One loop diagrams in a toy model.} 
\label{fig:oneloop_momentum}
\end{figure}

We derive Eq.~(\ref{eq:toy_scattering_amplitude_1}) in Sec.~\ref{sec:Kondo_introduction} 
in the nonrelativistic limit 
(see Ref.~\cite{Yasui:2013xr} and Sec.~\ref{sec:QCD_Kondo_effect} for relativistic computations). 
When the heavy impurity has a finite mass $M$ and a momentum $\vec{P}$,
the loop integral is given by (see Fig.~\ref{fig:oneloop_momentum}) 
\begin{widetext}
\begin{eqnarray}
M^{(1)}_{kl,ij}
&=&
G^{2} \T^\ac_{kl,ij} \int_{|\vec{k}\,| \ge k_{\mathrm{F}}} \cfrac{{\mathrm d}^{3}\vec{k}}{(2\pi)^{3}}
\cfrac{1}{\cfrac{1}{2m}|\vec{p}\,|^{2} + \cfrac{1}{2M}|\vec{P}\,|^{2} - \cfrac{1}{2m}|\vec{k}\,|^{2} + \cfrac{1}{2M}|\vec{P}+\vec{p}-\vec{k}\,|^{2} + i\varepsilon} \nonumber \\
&& 
+
G^{2} \T^\bc_{kl,ij} \int_{|\vec{k}\,| \le k_{\mathrm{F}}} \cfrac{{\mathrm d}^{3}\vec{k}}{(2\pi)^{3}}
\cfrac{1}{\cfrac{1}{2m}|\vec{p}\,^{\prime}|^{2} - \cfrac{1}{2M}|\vec{P}\,|^{2} - \cfrac{1}{2m}|\vec{k}\,|^{2} + \cfrac{1}{2M}|\vec{P}-\vec{p}\,^{\prime}+\vec{k}\,|^{2} - i\varepsilon},
\label{eq:amplitude_M1}
\end{eqnarray}
\end{widetext}
where the initial and final light quarks with mass $m$ have the three-dimensional momenta $\vec{p}$ and $\vec{p}\,^{\prime}$ on the Fermi surface ($|\vec{p}\,|=|\vec{p}\,^{\prime}|=k_{\mathrm{F}}$),
and $\vec{k}$ is the three-dimensional momentum in the loop integrals (cf.~Ref.~\cite{Yasui:2013xr}).
The first integral on the right-hand side gives the contribution from quarks above the Fermi surface, and the second one gives that from holes below the Fermi surface. When the heavy impurity mass $M$ is sufficiently large, the terms proportional to $1/M$ are neglected, 
and thus Eq.~(\ref{eq:amplitude_M1}) is reduced to
\begin{eqnarray}
\hspace{-1em}
M^{(1)}_{kl,ij}
&=&
G^{2} \T^\ac_{kl,ij} \int_{|\vec{k}\,| \ge k_{\mathrm{F}}} \cfrac{{\mathrm d}^{3}\vec{k}}{(2\pi)^{3}}
\cfrac{1}{\epsilon_{\mathrm{F}} - \cfrac{1}{2m}|\vec{k}\,|^{2} + i\varepsilon} 
\label{eq:amplitude_M2}
\\
&& 
+
G^{2} \T^\bc_{kl,ij} \int_{|\vec{k}\,| \le k_{\mathrm{F}}} \cfrac{{\mathrm d}^{3}\vec{k}}{(2\pi)^{3}}
\cfrac{1}{\epsilon_{\mathrm{F}} - \cfrac{1}{2m}|\vec{k}\,|^{2} - i\varepsilon},
\nonumber 
\end{eqnarray}
with the Fermi energy $\epsilon_{\mathrm{F}}=k_{\mathrm{F}}^2/2m$.
By using the absolute value of the energy measured from the Fermi surface,
\begin{eqnarray}
E = \cfrac{1}{2m} |\vec{k}\,|^{2} - \epsilon_{\mathrm{F}}
,
\end{eqnarray}
for particles ($|\vec{k}\,| \ge k_{\mathrm{F}}$), and
\begin{eqnarray}
E = -\cfrac{1}{2m} |\vec{k}\,|^{2} + \epsilon_{\mathrm{F}}
,
\end{eqnarray}
for holes ($|\vec{k}\,| \le k_{\mathrm{F}}$),
Eq.~(\ref{eq:amplitude_M2}) is rewritten as
\begin{eqnarray}
\hspace{-2em}
M^{(1)}_{kl,ij}
&=&
G^{2} \T^\ac_{kl,ij} \int_0^\infty
\cfrac{\rho(E)}{-E + i\varepsilon}  {\mathrm d}E 
\nonumber 
\\
&& \hspace{0.3cm} 
+
G^{2} \T^\bc_{kl,ij} \int_0^\infty \cfrac{\rho(E)}{E - i\varepsilon} {\mathrm d}E,
\label{eq:amplitude_M3}
\end{eqnarray}
with a density of states 
$\rho(E)=(2m)^{3/2}\sqrt{E+\epsilon_{\rm F}}/(4\pi^2).$ 
As a result, we obtain Eq.~(\ref{eq:toy_scattering_amplitude_1}).
We can approximate $\rho(E)$ as a constant 
$\rho_0=(2m)^{3/2}\sqrt{\epsilon_{\rm F}}/(4\pi^2)$ for small $E \ll \epsilon_{\mathrm{F}}$.

\section{Integral formulas}
\label{sec:integral_formula}

The integral appearing in the loop calculations can be carried out as 
\begin{eqnarray}
 \int_{-1}^{1} h(t) \, {\mathrm d}t &=& \frac{1}{3} \log \frac{8}{\lambda} 
,
\end{eqnarray}
where 
\begin{eqnarray}
h(t) = \frac{1}{2(1-t) + \lambda/\sqrt{2(1-t)}} \, .
\end{eqnarray}
To examine the infrared singularity arising near the Fermi surface, 
we may expand the following integrals with respect to $|\lambda| \ll 1$ as 
\begin{eqnarray}
&& 
\int_{-1}^{1} t^2 \,  h (t) \, {\mathrm d}t 
= -1 + \frac{1}{3} \log \frac{8}{\lambda} + \cdots, 
\\
&& 
 \int_{-1}^{1} (1-t^2) \,  h (t) \,{\mathrm d}t 
= 1 - \frac{2\pi}{3\sqrt{3}} \lambda^{2/3} + \cdots
,
\end{eqnarray}
and 
\begin{eqnarray}
&&
 \int_{-1}^{1}   h^2 (t) \, {\mathrm d}t 
= \frac{2\pi}{9\sqrt{3}} \lambda^{-2/3} + \frac{1}{8} + \cdots
, 
\\
&& 
 \int_{-1}^{1} t^{2}\,  h^2 (t) \, {\mathrm d}t 
= \frac{2\pi}{9\sqrt{3}} \lambda^{-2/3} + \frac{17}{24} - \frac{1}{3} \log \frac{8}{\lambda} + \cdots
,
\nonumber
\\
\\
&& 
 \int_{-1}^{1} (1-t^{2}) \,  h^2 (t) \, {\mathrm d}t 
= -\frac{5}{6} + \frac{1}{3} \log \frac{8}{\lambda} + \cdots
.
\end{eqnarray}

\section{Expansion by a small velocity}
\label{sec:v}

Substituting the kinematics specified in the beginning of Sec.~\ref{sec:solution}, 
the coefficient functions $F_{1}(q)$ and $F_{2}(q)$ 
in Eqs.~(\ref{eq:function_F1}) and (\ref{eq:function_F2}) are rewritten by 
\begin{eqnarray}
\hspace{-1em}
F_{1}(q) &=& \int_{-1}^{1} {\mathrm d}t 
\frac{1}{v_{0}(k_{\mathrm F}-q) - k_{\mathrm F} v \cos \alpha + vq t + i\varepsilon}, \\
\hspace{-1em}
F_{2}(q) &=& \int_{-1}^{1} {\mathrm d}t 
\frac{t}{v_{0}(k_{\mathrm F}-q) - k_{\mathrm F} v \cos \alpha + vq t + i\varepsilon}.
\end{eqnarray}
After some calculation, we obtain $F_{1}(q)$ as 
\begin{eqnarray}
F_{1}(q) &=& 
\cfrac{1}{vq} \Bigg( \log \cfrac{\sqrt{1+v^2}-v}{\sqrt{1+v^2}+v} 
\\
&&
+ \log \cfrac{q-(\sqrt{1+v^2}+v)(\sqrt{1+v^2}-v \cos \alpha)k_{\mathrm F}}{q-(\sqrt{1+v^2}-v)(\sqrt{1+v^2}-v \cos \alpha)k_{\mathrm F}} \Bigg) 
\nonumber
, 
\end{eqnarray}
and $F_{2}(q)$ as 
\begin{eqnarray}
\hspace{-0.5cm}
F_{2}(q) = \cfrac{1}{vq} \Big\{ 2 - \left( v_{0}(k_{\mathrm F}-q) - vq \cos \alpha \right) F_{1}(q) \Big\}
,
\end{eqnarray}
when $ \left| \sqrt{1+v^2}(q-k_{\mathrm F}) + vk_{\mathrm F} \cos \alpha \right| \ge vq$, because the energy scale $\Lambda_{0}$ is 
the starting energy scale in the RG equation. 
We find series representation of $F_{1}(q)$ and $F_{2}(q)$ at a small velocity $v\ll1$ as 
\begin{eqnarray}
F_{1}(q\!=\!k_{\mathrm F}+\Lambda) &=& 
-\cfrac{2}{\Lambda} + \cfrac{2k_{\mathrm F}\cos \alpha}{\Lambda^{2}} v + {\cal O}(v^{2}), 
\label{eq:F1+}
\\
F_{2}(q\!=\!k_{\mathrm F}+\Lambda) &=& \cfrac{2(k_{\mathrm F}+\Lambda)}{3\Lambda^{2}} v + {\cal O}(v^{2}),
\end{eqnarray}
when the light-quark momentum $q$ is larger than the Fermi momentum, 
and 
\begin{eqnarray}
F_{1}(q\!=\!k_{\mathrm F}-\Lambda) &=& \cfrac{2}{\Lambda} + \cfrac{2k_{\mathrm F}\cos \alpha}{\Lambda^{2}} v + {\cal O}(v^{2}), \\
F_{2}(q\!=\!k_{\mathrm F}-\Lambda) &=& 
\cfrac{2(k_{\mathrm F}-\Lambda)}{3\Lambda^{2}} v + {\cal O}(v^{2})
\label{eq:F2-},
\end{eqnarray}
when the light-quark momentum $q$ is smaller than the Fermi momentum. 


\begin{thebibliography}{99}

\bibitem{Alford:2007xm}
 M.G.~Alford, et al. 
 Rev.\ Mod.\ Phys.\  {\bf 80}, 1455 (2008)
 [arXiv:0709.4635 [hep-ph]].

\bibitem{1964PThPh..32...37K}
 J.~Kondo,
 Prog.\ Theor.\ Phys.\  {\bf 32}, 37 (1964).


\bibitem{1997kphf.book.....H}
 A.C.~Hewson,
 {\it The Kondo Problem to Heavy Fermions}
 (Cambridge University Press, 1997).


\bibitem{Abrikosov1965}
 A.A.~Abrikosov,
 ``Electron scattering on magnetic impurities in metals and anomalous resistivity effects,''
 Physics  {\bf 2}, 5 (1965).





\bibitem{0022-3719-3-12-008}
 P.W.~Anderson,
 J. of Phys. C: Solid State Physics {\bf 3}, 2436 (1970).

\bibitem{RevModPhys.47.773}
 K.G.~Wilson,
 Rev.\ Mod.\ Phys.\ {\bf 47}, 773 (1975).

\bibitem{Yamada}
 K.~Yamada,
 {\it Electron Correlation in Metals} (Cambridge University Press, 2004).

\bibitem{Yasui:2013xr}
 S.~Yasui and K.~Sudoh,
 Phys.\ Rev.\ C {\bf 88}, 015201 (2013)
 [arXiv:1301.6830 [hep-ph]].





\bibitem{Isgur:1989vq}
 N.~Isgur and M.B.~Wise,
 Phys.\ Lett.\ B {\bf 232}, 113 (1989).

\bibitem{Isgur:1989ed}
 N.~Isgur and M.B.~Wise,
 Phys.\ Lett.\ B {\bf 237}, 527 (1990).

\bibitem{Isgur:1991wq}
 N.~Isgur and M.B.~Wise,
 Phys.\ Rev.\ Lett.\  {\bf 66}, 1130 (1991).

\bibitem{Manohar:2000dt}
 A.V.~Manohar and M.B.~Wise,
 Camb.\ Monogr.\ Part.\ Phys.\ Nucl.\ Phys.\ Cosmol.\  {\bf 10}, 1 (2000).

\bibitem{Neubert:1993mb}
 M.~Neubert,
 Phys.\ Rept.\  {\bf 245}, 259 (1994)
 [hep-ph/9306320].

\bibitem{Baym:1990uj}
 G.~Baym, et al. 
 Phys.\ Rev.\ Lett.\  {\bf 64}, 1867 (1990).

\bibitem{Son:1998uk}
 D.T.~Son,
 Phys.\ Rev.\ D {\bf 59}, 094019 (1999)
 [hep-ph/9812287].

\bibitem{Hsu:1999mp}
 S.D.H.~Hsu and M.~Schwetz,
 Nucl.\ Phys.\ B {\bf 572}, 211 (2000)
 [hep-ph/9908310].

\bibitem{Fukushima:2010bq}
 K.~Fukushima and T.~Hatsuda,
 Rept.\ Prog.\ Phys.\  {\bf 74}, 014001 (2011)
 [arXiv:1005.4814 [hep-ph]].

\bibitem{Evans:1998ek}
 N.J.~Evans, S.D.H.~Hsu and M.~Schwetz,
 Nucl.\ Phys.\ B {\bf 551}, 275 (1999)
 [hep-ph/9808444].

\bibitem{Kaiser2002255}
 N.~Kaiser, S.~Fritsch and W.~Weise,
 Nucl.\ Phys.\ A {\bf 697}, 255 (2002)
 [nucl-th/0105057].





\bibitem{PhysRev.147.223}
 K.~Yosida,
 Phys.\ Rev.\ {\bf 147}, 223 (1966).

\bibitem{PR} R.~D.~Pisarski and D.~H.~Rischke,
  Nucl.\ Phys.\ A {\bf 702}, 177 (2002)
  [nucl-th/0111070].
  
\bibitem{GR} A.~Gerhold and A.~Rebhan,
  Phys.\ Rev.\ D {\bf 68}, 011502 (2003)
  [hep-ph/0305108].
  
\bibitem{HWR} D.~F.~Hou, Q.~Wang and D.~H.~Rischke,
  Phys.\ Rev.\ D {\bf 69}, 071501 (2004)
  [hep-ph/0401152].
  
  
   \bibitem{Schaefer2}  T.~Sch\"afer,
  Nucl.\ Phys.\ A {\bf 728}, 251 (2003)
  [hep-ph/0307074].
  
\bibitem{Schaefer}
 T.~Sch\"afer and K.~Schwenzer, Phys.\ Rev.\ D{\bf 70}, 054007 (2004).
 

\end{thebibliography}
\end{document}